\documentclass[useAMS,usenatbib]{mn2e}
\usepackage{graphicx}
\usepackage{url}
\usepackage{epstopdf}
\usepackage{color}
\usepackage{bm}
\newcommand{\apj}{Astrophys.~J.}

\newcommand{\apjl}{ApJ Lett.}

\newcommand{\mnras}{MNRAS}

\newcommand{\pasp}{PASP}

\newcommand{\pasj}{PASJ}
\newcommand{\aap}{A\&A}
\newcommand{\aj}{A J}
\newcommand{\nat}{Nature}

\title[Small-scale Alignment Effects.]{Drag-Gravity torques on galaxies in clusters: radial small-scale
alignment effects.}

\author[Alejandro Gonz\'{a}lez S. \&  Lu\'{\a i}s F. A. Teodoro ]{Alejandro Gonz\'{a}lez S.$^{1,\,2}$\,and Lu\'{\a i}s F. A. Teodoro$^3$\\
$^1$Centro de Investigaci\'{o}n, Divisi\'{o}n Acad\'{e}mica de Ciencias B\'{a}sicas - Universidad Ju\'{a}rez Aut\'{o}noma de Tabasco, 
Tabasco, M\'exico\\
$^2$Unidad Acad\'{e}mica de F\'{\a i}sica, Universidad Aut\'{o}noma de Zacatecas, M\'{e}xico; \,\tt{a.gonzales@physics.gla.ac.uk} \\
$^3$Department of Physics and Astronomy, University of Glasgow, Glasgow, G12 8QQ, Scotland, UK;\,\tt{luis@astro.gla.ac.uk}}

\begin{document}

\date{\today}
\pagerange{\pageref{firstpage}--\pageref{lastpage}} \pubyear{2008}
%\maketitle

\label{firstpage}

\maketitle

\begin{abstract}
We calculate the torque on galaxies in clusters due to gravity and to dynamical friction forces in order to study the possible origin of small-scale alignment effects as the result of interactions with their environment. The equation of motion for the position angle of a galaxy is derived by using a ple model. We find that weak radial alignment effects can be produced by this mechanism involving only the most massive galaxies. We also introduce a dependence on the cluster eccentricity to our equations in order to explore the alignment of galaxies with the cluster's major axis. We find that in the inner regions of high eccentricity clusters, alignments of massive galaxies with the cluster's major axis dominate over the radial ones. This mechanism could account for the observed alignment effects of the most massive
%first ranked 
galaxies with the major axis of their host cluster. Our results suggest that dynamical friction is a viable generator of alignment only for the most massive cluster galaxies. For the observed alignments of normal galaxies a primordial origin has to be explored.
\end{abstract}
\begin{keywords}
%Galaxies: alignments, Clusters: formation, Galaxies: evolution
Galaxies: evolution - Galaxies: formation - Galaxies: clusters: general
\end{keywords}

\section{Introduction}

The observational detection of different kinds and degrees of non-random orientation effects of galaxies and clusters has a long and rich history which goes back in time to the end of the 19th century (see Djorgovski 1983 for a review). Their prevailing importance lies in the fact that they naturally arise in various cosmological models of structure formation, with more preeminence in models with a flatter power spectra on large-scales
%{\textcolor{green}{being more consistent with those with flatter power spectra on large-scales}} 
\citep{West:1989ApJ...336...46W,Dekel:1988lsmu.book..561D}.  In these models, galaxy formation occurs in collapsing flattened structures where colliding density shock waves will tend to produce spin vectors aligned with the protocluster plane. In fact, \cite{Doroshkevich:1970Afz.....6..581D} predicted the existence of anisotropic orientation of galaxies as evidence for structure formation driven by a neutrino-dominated universe. He showed that the average bound region around a high density peak in the initial Gaussian density field is an ellipsoid with principal axes parallel with those of the deformation tensor at the peak within the formed filamentary structure. At the scale of  galaxy clusters, the ``peak-patch" model seems to validate such predictions in the context of cold dark matter universes \citep{Bond:1996Natur.380..603B}.  It is also very important to understand how alignment effects encode the influence of the environment on the formation and evolution of galaxies, but also how they contaminate the weak lensing measurements. 

On large scales, \cite{Gregory:1981ApJ...243..411G} showed that the major axis of the Perseus-Pisces supercluster coincides with the peak of the distribution of the position angle of galaxies. \cite{Djorgovski:1983ApJ...274L...7D} found the same effect for the Coma cluster and similar results have been reported for the Local supercluster \citep{MacGillivray:1985A&A...145..269M,Flin:1986MNRAS.222..525F,Kashikawa:1992PASJ...44..493K}. Furthermore, the evidence, for alignment of clusters' major axes over tens of megaparsecs \citep{Joeveer:1978MNRAS.185..357J,Binggeli:1982A&A...107..338B,West:1989ApJ...347..610W}
%Joever, Einasto and Tago 1978; Binggeli 1982; West 1989; Plionis 1993) 
could reinforce the predictions of alignment effects for flatter spectra on large scales.

On small-scales, \cite{Fong:1990MNRAS.242..146F} found some statistical evidence for three types of alignment: {\it{i}}) between neighbouring galaxies in a cluster; {\it{ii}}) between the major-axes of galaxies and the cluster's major-axis; and {\it{iii}}) between the position angle of a galaxy and the radial vector to the cluster centre. In this paper we are interested in the alignments of galaxies within clusters such as those reported by \cite{Hawley:1975AJ.....80..477H}, \cite{Thompson:1976ApJ...209...22T} and \cite{Adam:1980ApJ...238..445A}. More recently, this effect has also been shown %supported 
by \cite{Pereira:2005ApJ...627L..21P},  \cite{Agustsson:2006ApJ...644L..25A}, \cite{Torlina:2007ApJ...660L..97T} and \cite{Pereira:2008ApJ...672..825P}.  They found statistical evidence that galaxies appear to be aligned radially with respect to the cluster centre. Moreover, another point of interest concerns the evidence that in linear clusters the most massive galaxies  present the same orientation as their parent structures \citep{Sastry:1968PASP...80..252S,Carter:1980MNRAS.191..325C,Binggeli:1982A&A...107..338B,Trevese:1992AJ....104..935T}.
One could then ask whether such small-scale alignments are of primordial origin or have arisen as the result of the environmental influence on galaxies residing in clusters. In the latter case, primordial tidal interactions would seem an obvious candidate to produce, or even to suppress  any signal of anisotropy in the orientation \cite[e.g.][]{Binney:1979MNRAS.188..273B,Wesson:1982VA.....26..225W,Dekel:1985ApJ...298..461D}.

In this paper, we use a simple analytical model to investigate whether gravity and dynamical friction torques can account for stable orientations of galaxies as an evolutionary process. %In section \ref{sec:gravity}, we estimate the relative importance of the dynamical friction and the gravity, and calculate the timescales relevant for the existence of alignment effects. 
By taking into account the drag-gravity torque produced by the central part of the cluster, we derive in section \ref{sec:radialalignments}  the equation of motion for the position angle of galaxies. We show that for mass galaxies of  $\left[  10^{12}, 10^{13}  \right ] M_\odot$ the amplitude of the oscillations decays in a Hubble time and discuss the viability of this mechanism to produce alignment effects. A dependence of the alignment timescale on the cluster eccentricity is introduced in section \ref{sec:alignementmajoraxis}, and the alignment of galaxies with the cluster's major axis is also studied. We estimate which of these kind of alignments is the dominant effect. A brief discussion about the implications of our results on the theories of structure formation is given in section \ref{sec:discussion}.

%%%%%%%%%%%%%%%%%%%%%%%%%%% Radial Alignements
\vspace{-0.6cm}
\section{Radial Alignments} \label{sec:radialalignments}

In this section we quantify the effect on the dynamics of a galaxy, of the forces of gravity and dynamical friction, as a function of its distance to the cluster centre. 
For the cluster potential we adopt the triaxial model of \cite{Richtone:1980ApJ...238..103R} given by
\begin{equation}
\Phi\left(\mathbf{r} \right) = \Phi_c \ln \left[ \left(x/a\right)^2+ \left(y/b\right)^2+\left(z/c\right)^2 +1 \right],
\label{eq:richstone}
\end{equation}
with $ \Phi_c = 4\pi G \rho_c \left[ 2/3\left(1/a^2+ 1/b^2+1/c^2 \right)\right]^{-1}, \label{eq:richstone2}$
where $a$, $b$ and $c$ denote the main axes length and  $\rho_c$ 
is the core density and $\mathbf{r}\equiv (x,y,z)$. The gravitational force  $\mathbf{F_g}$ exerted by the cluster on a galaxy of mass $M$ is
\begin{equation}
\mathbf{F}_g =  -M\nabla \Phi(\mathbf{r}) = - 2 M f(\mathbf{r}) a^2 \left( \frac{x}{a^2} \hat{\mathbf{x}}+ \frac{y}{b^2} \hat{\mathbf{y}}+
 \frac{z}{c^2} \hat{\mathbf{z}}  \right),
\label{eq:gravforce}
\end{equation}
where
\begin{equation}
f(\mathbf{r})  = \frac{\Phi_c}{a^2} \left  (  \frac{x^2}{a^2} + \frac{y^2}{b^2}+ \frac{z^2}{c^2} + 1 \right)^{-1},
\label{eq:fbarx}
\end{equation}
and $(\hat{\mathbf{x}},\hat{\mathbf{y}},\hat{\mathbf{z}})$ are the unit directions along $(x,y,z)$-axis, respectively.
In order to obtain a simple estimate of the relevant timescales let us assume the gravitational force due to a spherical cluster $(b=c=a)$. In this case, equation (\ref{eq:gravforce}) acquires the form
\begin{equation}
\mathbf{F}_g = -2 f(r)  M \mathbf{r}, %\equiv -q(r)M \mathbf{r},
\label{eq:gravforce1}
\end{equation}
and
\begin{equation}
f(r) = \frac{\Phi_c }{a^2}\left (\frac{r^2}{a^2}   +1 \right )^{-1}.
\label{eq:frspherical}
\end{equation} 

We shall now calculate the timescale for the position angle decay and relate it to the orbital and crossing times. %This is required in order to set the conditions under which anisotropic orientations are produced.
We define the time of alignment,  $t_{\scriptsize{alig}}$,  as the time necessary for the orientation of the galaxy to decay to a fraction $1/e$ of its initial value. The orientation will be specified by the angle formed between the major axes of the galaxy and the radius vector joining its centre of mass with the cluster centre, $\theta$. A further assumption is that the direction of the torque acting on a galaxy, due to the cluster potential, is roughly constant. The requirements for this premise to be valid are that the galaxy moves in quasi-radial orbit, and that  $t_{\scriptsize{alig}} \le t_{\scriptsize{orb}}$. Otherwise, the dynamics of the galaxy in the presence of a time-varying torque would be more complex, and N-body simulations would be required. This is beyond the scope of this work and will be considered elsewhere (Gonz\'{a}les and Teodoro, in preparation).  Suppose we find that  $t_{\scriptsize{alig}} \gg t_{\scriptsize{orb}}$  then no alignments would be observed other than those already present in the initial conditions. However, for alignments produced as an evolutionary process of the cluster, the number of galaxies displaying these non-random orientations must be very low under the premise that the initial distribution of position angles is random. On the other hand, if it is found that 
$t_{\scriptsize{alig}} \ll t_{\scriptsize{H}}$, an anisotropy in the orientations would be easily created. Whilst for  $t_{\scriptsize{alig}} \approx t_{\scriptsize{H}}$, there would be a diluted signal of alignments but far from an isotropic orientational distribution.

%%%%%%%% Dipole
     In our analysis we consider interactions with the mean cluster potential only, and we will neglect the interaction with neighbouring galaxies. For this external tidal field to modify the orientation of galaxies a nearly homogeneous distribution of surrounding galaxies would be required.
     The total torque experienced by a galaxy is the sum of the torque due to the drag force and to the gravitational force. We will derive the expressions for the torques acting on a galaxy, which we model in a manner analogous to \cite{Peebles:1969ApJ...155..393P} and \cite{Wesson:1982VA.....26..225W,Wesson:1984A&A...138..253W}. The approximation consists in associating a dipole to the structure of the galaxy. The dipole is formed by two masses  
 $m=M_{\mbox{\scriptsize{galaxy}}}/2$, whose separation is calculated from the position  $d = \xi R$, of the centre of mass of half a disk (or ellipsoid) of radius (or major axis) $R$ and uniform density. Thus  $\xi = 0.42$  for disk galaxies and $0.37$ for ellipsoidal ones.  An expression for the torque may be calculated for galaxies when the shape is specified by two unequal semi-axes. The force acting on each mass of the equivalent dipole, according to equation 
$\mathbf{F}_g^{(i)} (\mathbf{r}_i) = -2m f(\mathbf{r}_i) \mathbf{r}_i$, where  $\mathbf{r_1}$ and $\mathbf{r}_2$ are their position vectors. 
The small variation in the value of $f$ %(proportional to the density) 
at the position of the two masses is responsible for the torque.
%The difference in the forces acting on the masses is needed in order to calculate the torque: %If the length of the dipole is much smaller than the distance to the cluster centre, the forces  $\mathbf{F} _g ^{(i)}$   can be considered parallel.  Thus, the difference of the forces acting on a galaxy is
$ \Delta \mathbf{F}_g\equiv \mathbf{F} _g^{(2)} - \mathbf{F} _g^{(1)} = -2m \left[ f(\mathbf{r}_2)  \mathbf{r}_2 -   f(\mathbf{r}_1)  \mathbf{r}_1   \right]$. 
Let  $2\mathbf{d} = \mathbf{r}_2 - \mathbf{r}_1$  be the length of the dipole and 
$2\mathbf{r}_{\mbox{\scriptsize{cm}}} = \mathbf{r}_2+\mathbf{r}_1  $ the galaxy centre of mass galaxy. The torque is then
$\mbox{\boldmath$\tau$} _g =  \mathbf{d}\times \Delta \mathbf{F}_g$,
where $\times$ denotes outer product.  We obtain a useful approximation for the torque as a function of the %position 
angle between $\mathbf{d}$  and $\mathbf{r}_{\mbox{\scriptsize{cm}}}$, by carrying out a Taylor expansion of $\Delta \mathbf{F}_g $ and retaining the first order term only $ -\Delta \mathbf{F}_g/4m   $ $ \approx $ $ f (\mathbf{r}_{\mbox{\scriptsize{cm}}}) \mathbf{d} +  \left(\mathbf{d} \cdot \nabla f \right)\mathbf{r}_{\mbox{\scriptsize{cm}}} $.  This leads to
\begin{equation}
\mbox{\boldmath{$\tau$}}_g =  -4m   \left(
\mathbf{d} \cdot \nabla f   \right)   \mathbf{d} \times       \mathbf{r}_{\mbox{\scriptsize{cm}}}.
\label{eq:torque1}
\end{equation}
Making use of the dipole length and of the fact that in a spherical potential $f$ is a function of radius only, we obtain
\begin{equation}
\mbox{\boldmath{$\tau$}}_g = -4md^2 | f'|  r \cos \theta\sin \theta \hat{\mathbf{e}}_\perp,
\label{eq:torque3}
 \end{equation}
where  $f'$  denotes derivative respect the radius and $ \hat{\mathbf{e}}_\perp \equiv \mathbf{r_1}\times \mathbf{r_2}/| \mathbf{r_1} \times \mathbf{r_2}| $.

%%%%%% Dynamical Friction
 In order to evaluate the dynamical friction component of the torque, we 
 calculate the dynamical friction force on a galaxy with a typical velocity $\sigma$, 
 %$\chi = 1$,  where the velocity is measured in units of the cluster velocity 
 %dispersion $\sigma$: $\chi = v/\sqrt{2} \sigma$. 
where $\sigma$ is the cluster velocity dispersion. The dynamical friction expression for such a galaxy can be rewritten in the form
$\mathbf{F}^{(i)}_d= - k m \rho \mathbf{v},
\label{eq:dyn_friction_force}$ where $k =  2\pi \ln \Lambda G^2 {m}\sigma^{-3} \left [ \mbox{erf}(1) -  2 e^{-1}  /\sqrt{\pi} \right]$ and $\ln \Lambda \approx (6 - 8.6)$ is the Coulomb logarithm for clusters. Thus, torque is
\begin{equation}
\mbox{\boldmath{$\tau$}} _d \equiv \mathbf{d} \times \Delta  \mathbf{F}_d = - k m \rho(r)  \mathbf{d} \times   \left( \mathbf{v}_2 -\mathbf{v}_1  \right).
\label{eq:dyfriction_pair}
\end{equation}
Note $\rho(r)$ must be different for each of the masses, but since the galaxy size is small compared with the distance to the cluster centre, it is reasonable to adopt the density at the position of the centre of mass. The velocities can be expressed in terms of the peculiar velocity in the cluster $\dot{\mathbf{r}}$ and the angular velocity  $\omega \equiv \dot{\theta}$ as $ \mathbf{v}_{1,2} = \dot \mathbf{r}  \mp \mathbf{d} \times \mbox{\boldmath{$\omega$}}$ which yields
\begin{equation}
\mbox{\boldmath{$\tau$}} _d = -2 k m \rho d^2 \mbox{\boldmath{$\omega$}}.
\label{eq:torque_drag}
\end{equation}
%The sum of the frictional terms acting on each mass is proportional to $m^2/2$ not to $M^2$, leading to a total dynamical friction which half of what it would be for a point like mass M.
%; it would be one half that of the total dynamical friction \citep[see for instance][]{Binney:1987gady.book.....B}. 
Strictly, the total dynamical friction is not a sum of drag forces experienced for each mass: both halves of the ``two-body" galaxy are dragged by both wakes. 

%%%%%%%%% Equation of motion
The equation of motion for the position angle is derived by expressing the total torque on a galaxy of characteristic radius $R$. In terms of its inertia tensor  $I= 2mR^2$  and its angular velocity  that is,
\begin{equation}
\left | \mbox{\boldmath{$\tau$}}    \right | = 2 m R^2 \ddot{\theta}.
\label{eq:torque_inertia}
\end{equation}
%Substituting in equation (19) we obtain
%\begin{equation}
%\ddot{\theta} + 2 f'(r) r (d/R)^2 \sin \theta \cos \theta = 0.
%\label{eq:diff_eq_torque}
%\end{equation}
%%%%%%%%%%%%%%%%%%%%%%%%%%%
%\subsection{Equation of Motion: Results}
%\label{subsec:equation_of_motion}
If we invoke the geometric factor  $\xi = d/R$, the equation describing the motion of the position angle of the galaxy is obtained from expressions (\ref{eq:torque3}), (\ref{eq:torque_drag}) and (\ref{eq:torque_inertia}):
\begin{equation}
\ddot{\theta} + k\rho(r)  \xi^2 \dot \theta + 2 |f'(r)|r \xi^2 \sin \theta \cos \theta  = 0.
\label{eq:equation_motion}
\end{equation}
A damped harmonic oscillator analogy is evident: the major axis of the galaxy can either rotate or oscillate. It rotates if it has sufficient energy $E_{\mbox{\scriptsize{crit}}} =  \left[ \dot \theta ^2 -  |f'(r)| r \xi^2 \cos \theta \right] /2 $,
to carry it past the vertical upright position, but oscillates otherwise. If the galaxy rotates, its angular velocity is minimum near the position perpendicular to the radial alignment and maximum when is parallel to it. As the motion is damped by the action of the frictional term, rotation can eventually be replaced by oscillations around the equilibrium point corresponding to the radial alignment. 
%Galaxies rotating in a retrograde sense to the direction of the angular velocity produced by torquing are more stable to the tidal field than galaxies with either rotation in the same direction or no rotation. This pattern of motion has been observed in numerical simulations addressed to study the environmental effects on the internal dynamics of bar-like galaxies in clusters \citep{Miller:1982ApJ...253...58M,Miller:1988ASSL..140..353M,Farouki:1981ApJ...243...32F}. 
\begin{figure}
\begin{center}
{\scalebox{0.44}{\includegraphics{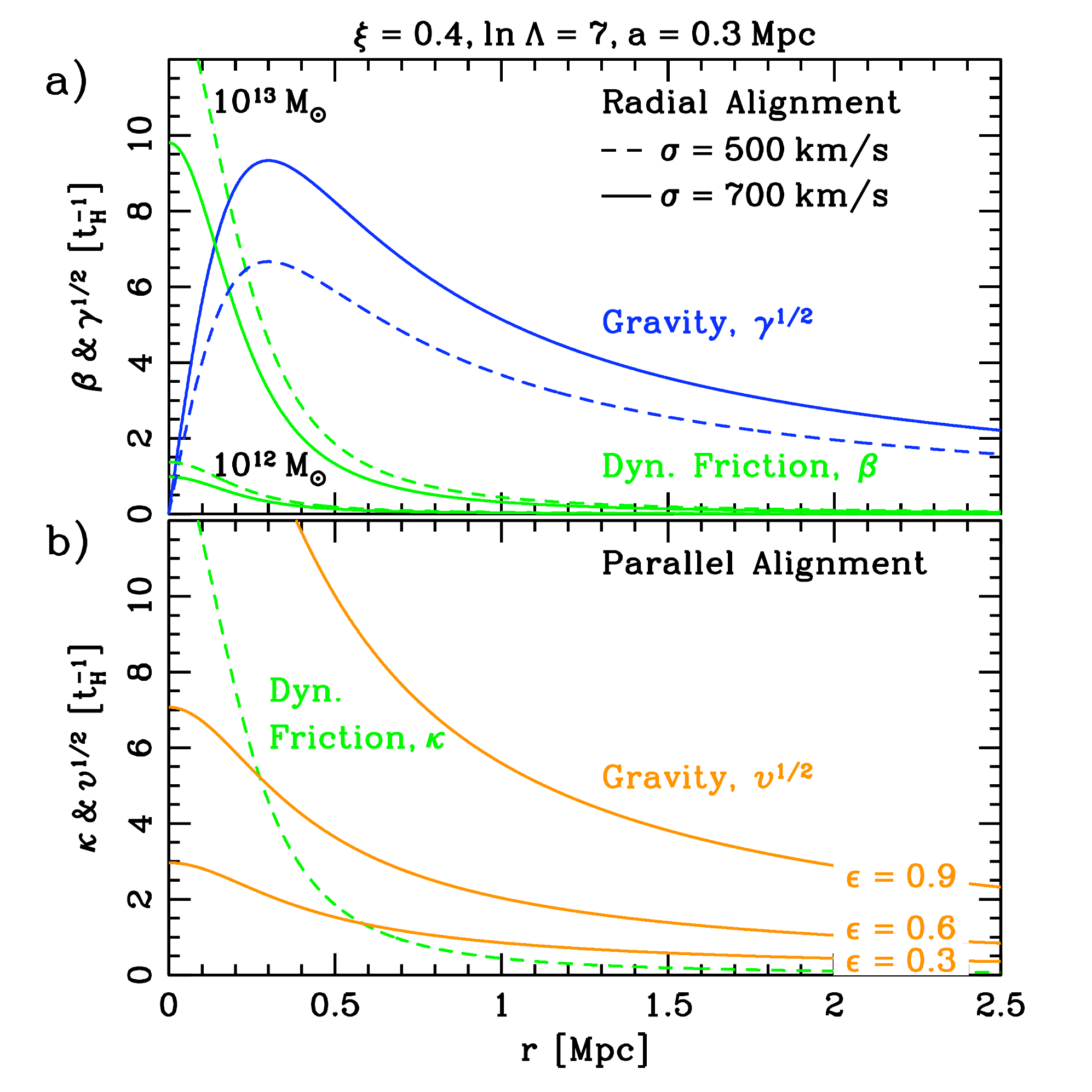}}}
%{\scalebox{0.43}{\includegraphics{fig1b.pdf}}}
\end{center}
\vspace{-.2cm}
\caption{a) Dynamical friction damping term compared with the gravitational term as a function of the galaxy mass, the cluster velocity dispersion and the distance to the cluster core; b) same as a) but here the gravitational term for  $\sigma = 500$\,km/s changes due to its dependence on the cluster eccentricity $\epsilon$. In the lower panel the Dynamical Friction term assumes $M = 10^{13}\,M_\odot$.}
\label{fig:radial_alignment}
\end{figure}

The Chandrasekahr's formula deduced under the assumption of interaction of point masses, only slightly overestimates the real friction that acts on an extended body \citep{Binney:1987gady.book.....B,Read:2006MNRAS.373.1451R}. Hence, the simple dipole model adopted in this study should be satisfactory, and a description in terms of the exact geometry of the galaxies would not produce an important change in the damping timescale.
The time required to damp the initial amplitude of the oscillations %(\ref{eq:equation_motion}), 
is contained in the parameters
\begin{equation}
\beta \equiv k\rho(r) \xi^2\quad \mbox{and} \quad \gamma \equiv 2 |f'(r)| r  \xi^2.
\label{eqs:oscillation_parameters}
\end{equation} 
%For homogeneous clusters we can relate $\beta$  to the prior determination $p$ of this timescale, by simply noting that $\beta = p \xi^2$. Taking  $\xi \approx 0.4$, and $p \approx 0.16 t_H^{-1} $ where $p$ corresponds to galaxies of mass  $10^{12} - 10^{13}\,M_\odot$ one obtains $\beta \approx 0.16 t_H^{-1} $ which gives a timescale $ t_{\scriptsize{alig}} \approx \left[  4-6 \right] t_H$.  For low mass galaxies this time increases by a factor of $10$, in proportion to the mass.
  In order to visualize whether the dynamical friction produced in the central regions of an inhomogeneous cluster is strong enough to damp the oscillations in a Hubble time, we have compared the parameters $\beta$ and  $\gamma^{1/2}$ as a function of the radius. The upper panel of figure\,(\ref{fig:radial_alignment}) shows such a comparison, the assumed parameters are: $\ln \Lambda = 7$, $\xi = 0.4$\,and $a=0.3$\,Mpc.
   
The gravitational term dominates in the outer regions of the cluster, where galaxies oscillate around the point of a perfect radial alignment. As the galaxy reaches the intermediate zones, the dynamical friction becomes of the same order of magnitude as the gravity, tending to damp the oscillations. In the innermost regions of the cluster, near $600$ kpc ($\sim 2a$) away from the centre, the dynamical friction has an increasing effectiveness. At this stage the oscillations are strongly damped, in a fraction of a Hubble timescale.
%\textcolor{red}{Can we add something about accretion and the gas being blown by the motion of the galxy in the cluster.}

One conclusion derived from these estimates is that any alignment produced by this mechanism should be very weak, as a result of  $t_{\scriptsize{alig}} \approx  t_H \approx  t_{\scriptsize{orb}}$. Moreover, because the adopted values of the parameters were those associated with the galaxies that experienced a stronger damp, any radial alignment generated should preferentially involve the most massive galaxies in the cluster. This result is not surprising because the dynamical friction grows in proportion to the mass. The observational evidence of radial alignments however, shows no preference for the most massive galaxies \citep{Thompson:1976ApJ...209...22T}.  A viable way of distinguishing between a primordial or an evolutionary origin of alignments, is by looking at the mass and position of the galaxies displaying this effect. Alignments arising from an evolutionary process, should be detected in the intermediate and innermost regions of the clusters and involve only the most massive galaxies. If, at early times, the cluster had an isotropic distribution of orientations then, after a Hubble time the dispersion of the new distribution would be a factor $e^{-1}$ away from the perfect radial alignment, showing some statistical evidence of alignments. Furthermore, a search for alignment effects by using the external galaxies could dilute a positive signal because the isotropy in the orientations has not been significantly affected by the dynamical friction. On the other hand, if it is found that the effect extends to the outer cluster regions involving low mass galaxies, it will be quite suggestive of a possible primordial origin.

%%% CLUSTER'S MAJOR AXIS
\vspace{-0.6cm}
\section{Alignment with the Cluster Major axis}
\label{sec:alignementmajoraxis}

The study of this type of alignment is performed more conveniently in cartesian coordinates. The torque on the galaxy is
$%\begin{equation}
\mbox{\boldmath{$\tau$}} _g = \left( \mathbf{r}_2 - \mathbf{r}_1  \right) \times \Delta \mathbf{F}_g/2,
$ where the differential gravitational force across the galaxy is obtained from equation (\ref{eq:gravforce})
\begin{equation}
\Delta \mathbf{F}_g =  - 2 m f(\mathbf{r}) a^2 \left( \frac{x_2-x_1}{a^2} \hat{\mathbf{x}}+ \frac{y_2-y_1}{b^2} \hat{\mathbf{y}}+ \frac{z_2-z_1}{c^2} \hat{\mathbf{z}} \right).
\end{equation}
As in the previous section, because the difference in the density of the cluster on the position of the masses is not considerable, one can assume that the value of $f$ is approximately equal to the density on the centre of mass of the galaxy i.e.  $ f(\mathbf{r}) \approx f(\mathbf{r}_1) \approx f(\mathbf{r}_2) $. Furthermore, in order to obtain some insight on the general feature of the motion let us restrict  to the study the problem in the plane $x-y$ and assume that $b = c$. Hence, the  $\hat{\mathbf{z}}$  component of the torque is
\begin{equation}
\mbox{\boldmath$\tau$} _{g,z} = -m f(\mathbf{r}) a^2 \left\{  \left( x_2-x_1 \right) \left(y_2-y_1 \right) \left(  b^{-2} -  a^{-2} \right)   \right\}.
\end{equation}
Defining the eccentricity of the cluster as  $\epsilon^2 = 1 - (b/a)^2$, and $\phi$ 
to be the angle between the galaxy major axis and the cluster's major axis, we can express the gravitational torque as
$%\begin{equation}
\mbox{\boldmath$\tau$} _{g,z} = -m a^2 f(\mathbf{r}) d^2 \epsilon^2 b^{-2}\sin \phi \cos \phi \hat \mathbf{z}.
$ %\end{equation}
After the geometric factor is introduced by using the inertia tensor  $I = 2 m R^2$ and the dynamical friction added, we obtain
\begin{equation}
\ddot \phi + k \rho(\mathbf{r}) \xi^2 \dot{\phi} + 2 f(\mathbf{r})\xi^2 {\epsilon^2}{(1-\epsilon^2)}^{-1}\sin \phi \cos \phi = 0. 
\end{equation}
Note that this term vanishes for $\epsilon = 0$ because the alignment of galaxies with the ``major axes" of a spherical cluster loses its meaning. The bottom panel of  figure\,(\ref{fig:radial_alignment}) shows the comparison of the relative size of the parameters
\begin{equation}
\kappa \equiv k\rho(\mathbf{r})\xi^2 \quad \mbox{and} \quad \upsilon \equiv 2 f(\mathbf{r}) \xi^2 {\epsilon^2}{(1-\epsilon^2)}^{-1},
\end{equation}
for different cluster eccentricities. For massive galaxies in clusters of low velocity dispersion, the drag and gravity contributions are approximately of the same order of magnitude for the inner regions. The relevance of the term containing the cluster eccentricity is that it projects the importance of the gravitational interaction. Tidal interactions between more eccentric shapes are stronger than those experienced for symmetric structures. Thus, the gravitational term dominates the oscillations of the galaxies major axes for the flattest clusters. 
\begin{figure}
\begin{center}
{\scalebox{0.45}{\includegraphics{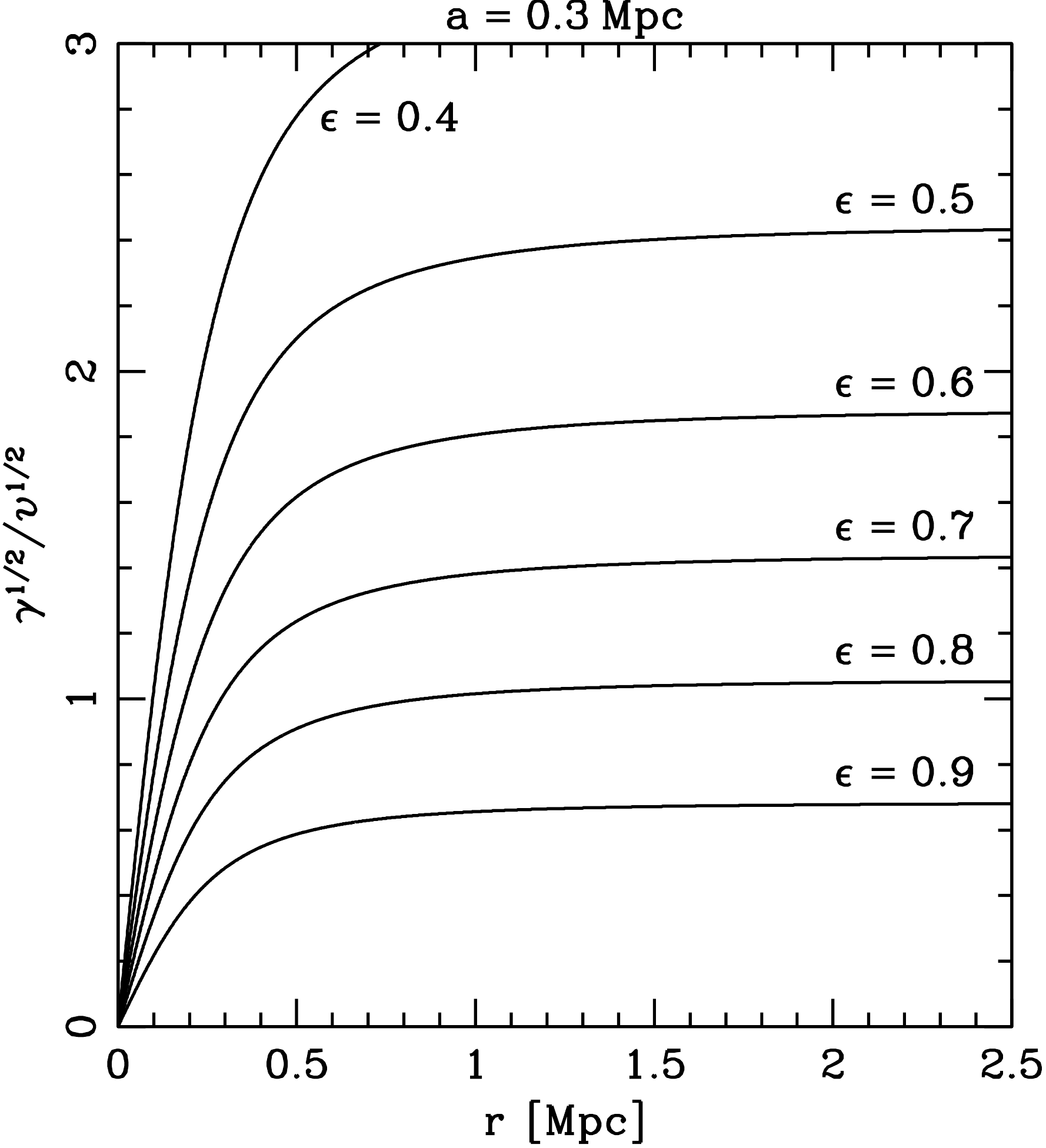}}}
\end{center}
\vspace{-.2cm}
\caption{Ratio of  $\gamma^{1/2}$  and $\upsilon^{1/2}$: the gravitational factor responsible for the radial alignment to that responsible for the parallel alignment for different cluster eccentricities. The asymmetry of the cluster potential dominates in the innermost region of the cluster $(\sim 2a)$. I}
\label{fig:gamma_v}
\end{figure}
However, the observed ellipticity of clusters (typically $\epsilon \approx 0.5$) lies within the interval $[0-0.6]$ %, or equivalently within the interval of eccentricities $[0-0.9]$, 
where the dynamical friction is comparable with the gravitational effect. One can infer then, that if such a mechanism operates in clusters of galaxies, in principle alignments other than those between the most massive galaxies with the major axes of the flattest clusters should be observed.  Because this mechanism does not damp the position angle oscillations for all the mass range of galaxies, one can conclude that alignment effects of this kind would hardly be due to an evolutionary process like the one considered in the present paper. In an asymmetric cluster there is an interplay between the radial and parallel alignment. A measure of which of these two effects dominates the final orientation of galaxies located in the innermost regions of clusters is achieved by comparing the parameters  $\gamma$ and $\upsilon$. From figure\,(\ref{fig:gamma_v}) we observe that in the outer regions of the cluster, the gravitational term responsible for the radial alignment is larger than that corresponding to the alignment with the cluster's major axis. In the inner regions this situation changes and the parallel alignment becomes the dominant one. In this region of the cluster, the dynamical friction grows to a size comparable with the gravitational term. Therefore, the galaxy tends to move toward the cluster centre and to get aligned with its major axis. For nearly symmetric clusters, the radial alignment always dominates over the parallel alignment but the dynamical friction is not significant, so that no alignments are present, except for galaxies of $10^{12}-10^{13} M_{\odot}$.

\vspace{-0.6cm}
\section{ DISCUSSION AND CONCLUSIONS}
\label{sec:discussion}

We have studied the possibility of obtaining non-random orientations of galaxies in clusters produced by the combination of gravitational tidal interactions with the cluster potential and dynamical friction.
     The orientation of non-rotating galaxies is unstable in the presence solely of gravitational tidal interactions, i.e., they oscillate around the position of equilibrium corresponding to the alignment between the main axes of the galaxy with the radial vector to the cluster centre, or parallel to the cluster's major axis. With the inclusion of the dynamical friction we found that these oscillations can be significantly damped only for galaxies with masses in excess of  $10^{12} M_\odot$. The damping occurs in a Hubble timescale, which is of the same order of magnitude as the orbit decay timescale. Thus, the final orientation of the galaxies that have time to reach the cluster centre will be determined by the eccentricity of the cluster. Whilst at the outer regions of nearly symmetric clusters the gravitational term responsible for the radial alignment is larger than that corresponding to the alignment with the clusters's major axis, in the inner regions the parallel alignment becomes the dominant one where the anisotropy of the tidal field is maximum. For clusters of low eccentricity,  $\epsilon \le 0.6$
(ellipticity  $ \le 0.3 $), parallel alignment is only important very close to the core. For flatter clusters, where the tidal field is markedly anisotropic, the gravitational term ($\upsilon$) is more important than it is in the symmetric case ($\gamma$). The anisotropy of the tidal field is of the same order of magnitude of the radial action of gravity even at larger distances of the cluster centre (see Figure\,\ref{fig:radial_alignment}). In these regions, the dynamical friction damps the oscillations around the cluster's major axis in a fraction of a Hubble time.
     These results makes plausible that massive galaxies that have had time to reach the cluster centre are aligned with the cluster's major axis. One can then conjecture that the final result of these encounters would produce a giant, central galaxy aligned with the cluster's major axis. Even when tidal disruption or merger with other galaxies may occur, the direction defined by the clusters' major axes would be preferential to the motion of the remnants of these processes. The tidal field defines a preferred direction, along the major axis for eccentric clusters, and along the radius vector to the cluster centre if it is nearly spherical. This mechanism could produce anisotropic mergers of objects with quasi-parallel vectors in the central region. Moreover, because of the low number of objects of very massive galaxies in a cluster, mergers would be highly anisotropic and the final product would have a non-neglegible peculiar velocity. Such central galaxies have been reported by \cite{Hill:1988ApJ...332L..23H}.
 
     There is some numerical evidence that the orientation of the first-ranked galaxies formed by the merging of sub-structure in clusters of galaxies, is determined by the initial ellipticity of the cluster \citep{Rhee:1990MNRAS.243..629R} and that possibly orientational memory is preserved by the galaxies. However, such a model shares the drawback of the merging theories which cannot account for the high peculiar velocities of cD galaxies. Further, the merging theory cannot account for alignment of galaxies with the clusters major axes in the external regions.
     On the other hand, with the mechanism of drag and gravity torques studied here, alignment of galaxies could not be produced at the outer regions of clusters where the damping timescale is several times the Hubble time. Thus, the preference for the most massive galaxies at the central region of clusters suggest that gravitational tides combined with dynamical friction are not the only viable generator of the observed alignments, which include low mass galaxy alignments extending to the outer region of the clusters. Radial and parallel alignment effects must therefore be compatible with the idea of a primordial origin. However, one of the reasons why this effect is not strongly observed in all the clusters is that there are many other environmental effects which could suppress the alignments at a post cluster formation era. The final orientation of galaxies would be difficult to determine for galaxies with non-radial orbits. These are some of the factors not contemplated in this work.
     Concerning a primordial origin,  \cite{West:1989ApJ...336...46W} have shown that alignment effects of clusters of galaxies naturally arise in Universes dominated by Mixed Dark Matter. In fact the small dispersion in cD's luminosities, for a number of possible cluster physical conditions suggest that cD progenitors should be special ab initio \citep{West:1994MNRAS.268...79W}. \cite{Barnes:1987ApJ...319..575B} and \cite{Bond:1987nngp.proc..388B} have also suggested that non-random orientation of density perturbations could be primordial, and be present at the initial conditions. Lambas, Groth and Peebles have found evidence of alignment of elliptical galaxies with the major axis of their host clusters, but not for spiral galaxies. If the orientation of galaxies in-deed reflect the primordial orientation of maxima of the density field, then the angular momentum of galaxies need not have any connection with the angular momentum of elliptical galaxies, which have a lower specific angular momentum. These ideas are being investigated (Gonz\'{a}lez \& Teodoro, in preparation).

LFAT acknowledges the financial support of the Leverhulme Trust. A G thanks COZCYT and SUTUAZ for support.

\vspace{-0.8cm}
%\bibliography{mnrasmnemonic,bibliography}
\bibliographystyle{astron}
%\bibliography{bibliography}

\pagestyle{empty}

\label{lastpage}
\end{document}